\documentclass[a4paper, 12pt]{article}
\usepackage{graphicx,graphics}
\usepackage{placeins}
\usepackage{url}
\usepackage[hidelinks]{hyperref}
\usepackage{amsmath}	
\usepackage[hidelinks]{hyperref}

\usepackage{tikz,xcolor,hyperref}
\definecolor{lime}{HTML}{A6CE39}
\DeclareRobustCommand{\orcidicon}{
	\begin{tikzpicture}
	\draw[lime, fill=lime] (0,0) 
	circle [radius=0.16] 
	node[white] {{\fontfamily{qag}\selectfont \tiny ID}};
	\draw[white, fill=white] (-0.0625,0.095) 
	circle [radius=0.007];
	\end{tikzpicture}
	\hspace{-2mm}
	}
	\foreach \x in {A, ..., Z}{\expandafter\xdef\csname orcid\x\endcsname{\noexpand\href{https://orcid.org/\csname orcidauthor\x\endcsname}
			{\noexpand\orcidicon}}
}

\begin{document}

\title{
The meaning of entropy
\\
(demonstration of a much needed theorem)
}
\author{Penha Maria Cardozo Dias \footnote{penha@if.ufrj.br} \orcidA{} 
\\
Instituto de F\'isica, Universidade Federal do Rio de Janeiro
\\ Rio de Janeiro, Brazil}

\date{}
\maketitle

\begin{abstract}

\noindent
The association of information with entropy has been argued on plausibility arguments involving the operation of imaginary engines and beings, and it is  not a universal theorem. In this paper, a theorem by Charles Bennett on reversible computation that associates entropy with erasure of information is recognized as this much needed theorem. It is proposed a real, non thermal engine, operated by humans. It is proved:
(1) The engine obeys two laws, identical {\it mutatis mutandis}  to the two laws of thermodynamics; therefore, the  entropy that arises in the operation of the engine has the same meaning of the entropy that arises in the operation of thermal engines. (2) The  engine operates in stages similar to the stages in Bennett's three tapes  reversible computer; therefore the entropy in the engine has the same meaning of the  entropy in 
computation. The conclusion is that also the thermal entropy  is a measure of erased or missing information. 
As a side result, information is measured in physical units, which complies with Landauer's principle. A prototype at work is shown in video.

\vspace{0.2cm}
\end{abstract}

\section{Introduction}
\label{intro}
In 1824, Nicolas L\'eonard Sadi Carnot \cite{Carnot} stated the categories of the theory of heat; among these categories, there is  the `` principle" ({\it Carnot's principle}) on which heat engines work, and a theorem ({\it Carnot's theorem}).  However, the two laws of thermodynamics were stated only in 1850  by Rudolf Julius Emmanuel Clausius \cite{Clausius1850}. He corrected both Carnot's principle and the theorem. These corrections led Clausius to energy conservation, and  
to postulate a new, qualitative law on the direction of heat flow. In 1854 \cite{Clausius1854}, he gave a quantitative, analytic  formulation  to the former  law on heat flow, and introduced a new quantity, the entropy. Clausius's deductions of the two laws consist in seeking conditions to restore the initial conditions of the working substance, and of the  reservoirs, respectively \cite{DiasUppsala,Dias1995}. 
Almost two hundred years later, the meaning of the second law is still being discussed, as well the meaning of the entropy, and the range of applicability of the law has been extended to apparently uncorrelated areas, such as computation \cite{LeffRex1,LeffRex2}. Once the laws are understood as restoring conditions, there is an analogy with the process of computation: ``restoring" is analogous to the logical operations ``restore the operation $p\wedge q$ ($p$ and $q$)"  and ``restore the operation $p\vee q$ ($p$ or $q$)" \cite{Landauer1961,BennettLandauer};  irreversibility of the logical operations ``and" and ``or" makes computation irreversible. However Charles Bennett \cite{Bennett1973} proved a theorem to the following effect: a computation can be made reversible, if the script is recorded on an extra tape (the garbage or history tape), which allows to trace the script back. 
A reversible computer is then imagined as having three tapes: a tape keeps record of all the calculations, and the other two respectively keep record of the input and output (Section~\ref{sec:thereversiblecomputer}). 
Reversibility is  associated with the capability to retrieve information from the history tape; accordingly, Bennett interprets irreversibility as due to {\it erasure of information}. 

The association of information with the capability to reverse a process dates back to Maxwell's Demon, a creature that acts on information or knowledge in order  to reverse a situation of thermodynamic equilibrium (Section~\ref{sec:epistemicentropy}). The arguments rely on thought gadgets, and on imaginary creatures to operate them \cite{LeffRex1,LeffRex2}. As convincing as the arguments are, they do not have the strength of a universal theorem, and only show that the association of entropy with information is highly plausible, appealing and desirable. 

In this paper, Bennett's theorem is recognized as the needed universal theorem that associates the  thermodynamic entropy with information or knowledge (epistemic meaning). 
If it is possible to make a 1-1 correspondence 
between the operation of Bennett's  
three-tapes 
 computer  and the operation of
Carnot's thermodynamic engine, then (ir)reversibility  in thermodynamics follows from the (im)possibility to recover information; the thermodynamic entropy also is  a measure of erased information.
The proof  relies on the operation of an engine. However, the engine is real: a prototype has been built, its functioning tested, and a video has been prepared; it dispenses with creatures from the Hades (the operator is an {\em Homo Sapiens Sapiens}). 

The engine is  a metaphor of a thermal engine, in which the part of heat is played by 
air (Section~\ref{sec:engine}).
Closely following  Clausius's demonstrations of the two laws of thermodynamics  (Section~\ref{sec:clausiustwooperations}), it is shown that the engine obeys identical laws  (Section~\ref{sec:twolaws}); this  generalizes the laws of thermodynamics to a generic engine, henceforth called 
{\it un-thermal dynamic engine}.\ignorespaces 
\footnote{The air is similar to ``caloric".  Perhaps this is  not mere coincidence. Sanborn Brown \cite{SanbornBrown} shows that the caloric theory was successful in building thermodynamic concepts, such as heat capacity, sensible and latent heats, equilibrium of temperature,  quantity of heat and temperature.}
There is a correspondence 1-1  between  the engine and the three-tapes computer  (Section~\ref{sec:theunthermalcomputer}).
Finally,  
information is measured  in units of quantity of air.


\section{The un-thermal dynamic engine}
\label{sec:engine}
\begin{figure}[!!!h]
\centering
\includegraphics[width=11cm]{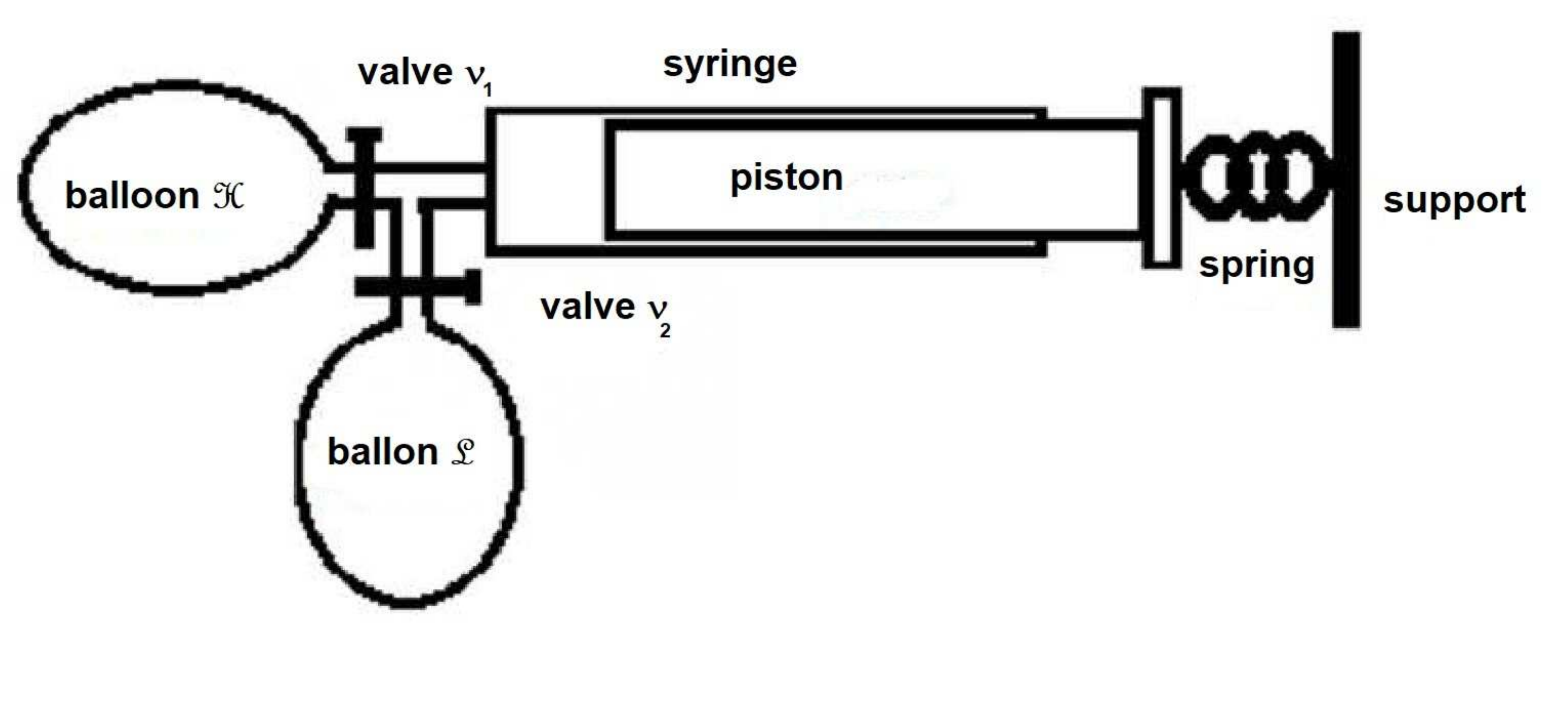}
\caption{The un-thermal dynamic engine. A syringe of cross section $A$ has its piston attached to a spring of relaxed length $L$, fixed to a support. $\cal{H}$  and $\cal{L}$  are balloons. $\nu_1$ and $\nu_2$ are valves that let air in or out the balloons. Graduations on the syringe give the distance moved by the piston or the distance the spring either contracts or expands}
\label{fig:1}
\end{figure}

 Figure~\ref{fig:1} is the blueprint of the 
un-thermal dynamic engine.  The direct cycle is described in  Section~\ref{sec:directcycle}, and the reversed cycle in Section~\ref{sec:reversedcycle}.    

\begin{figure}[!!!h]
\centering
\includegraphics[width=10cm]{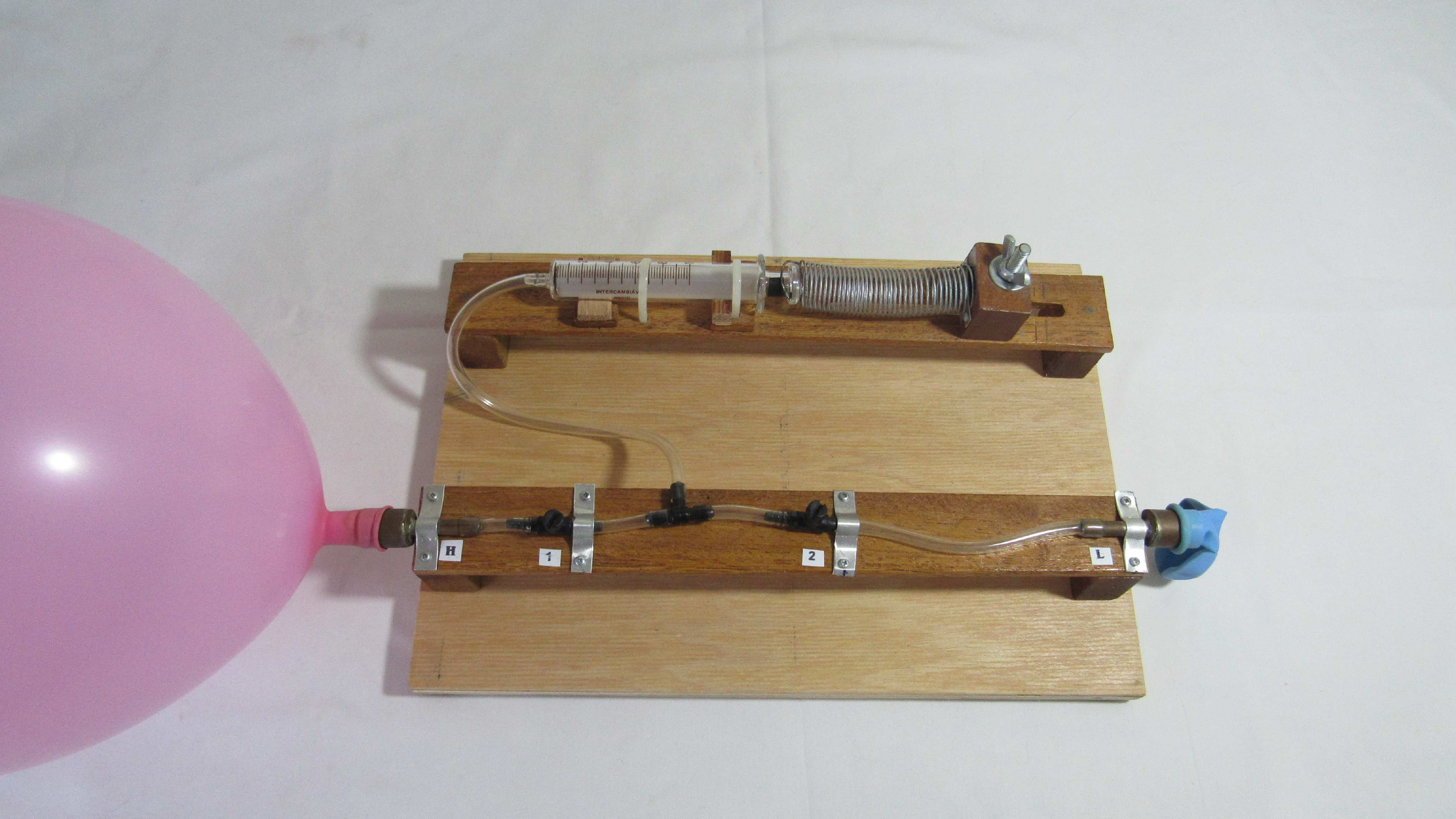}
\caption{The prototype. The wooden base has 
$32~\mathrm{cm}\times 22~\mathrm{cm}$. $\cal{H}$ (full) is on the left, $\cal{L}$ (empty) is on the right. The syringe is made of glass, and it is linked by narrow plastic tubes to the balloons.}
\label{fig:prototype}
\end{figure}

Figure~\ref{fig:prototype} shows a prototype of the engine. Its operation is shown in a video: 
\begin{center}
\url{https://youtu.be/HwMv-Rg2oTQ}
\end{center}

Table~\ref{tab:dictionary} is a dictionary that  translates concepts of the un-thermal dynamic engine into concepts of Carnot's thermal engine.
 The engine is supposed to be in a heat-bath, the environment, so that  the   state of the spring is defined by the pressure 
and the ``volume". The length of the spring is 
$0\leq x \leq L$, $k$ is the spring constant,  and $A$ the cross section of the syringe. The engine undergoes quasi-static processes (in the sense of Table~\ref{tab:dictionary}), and 
 no transfer of air happens without  involving work, a demand made by Carnot \cite{Carnot}.
{
\renewcommand{\baselinestretch}{0.8}
\begin{table}[!h]
\caption{Translation dictionary}
\centering
\label{tab:dictionary} 
\begin{tabular}{l|l}
\\
{\bf un-thermal dynamic engine} & {\bf  thermodynamic engine}
\\
\hline\hline
	 {\footnotesize air} &  {\footnotesize heat}
\\
\hline
 {\footnotesize balloon $\cal{H}$} &  {\footnotesize hot reservoir}
\\
\hline
		 {\footnotesize balloon $\cal{L}$} &  {\footnotesize cold reservoir}
\\
\hline
	 {\footnotesize work done by  the air on or by the spring} &
 {\footnotesize transferred heat} 
\\
\hline
  {\footnotesize spring}& {\footnotesize working substance}
\\
\hline
 {\footnotesize potential energy of the spring}&  {\footnotesize internal energy} 
\\
\hline
{\footnotesize air and piston  always in contact:}
&
\\
{\footnotesize
air and spring expand or compress}
& {\footnotesize quasi-static process}
\\
{\footnotesize through states of equilibrium}
&
\\
\hline
  {\footnotesize 
work is done by (or on) the air 
in the syringe 
}
	& {\footnotesize isothermal process}
\\
 {\footnotesize 
{\bf with} transfer air from (or to) a balloon}&
\\
\hline
  {\footnotesize  work is done by (or on) the air 
in the syringe 
} 
	& {\footnotesize adiabatic process}
\\
 {\footnotesize {\bf without}  transfer of air from (or to) a balloon}
\\
\hline
{\footnotesize ``volume" of spring: $V=Ax$}
&
{\footnotesize volume occupied by the gas}
\\
\hline
{\footnotesize pressure of the spring: $P_{x}=\frac{k}{A}x$}
&{\footnotesize pressure of the gas}
\\
\hline
{\footnotesize  $P=-\frac{k}{A^2}V$}&{\footnotesize equation of state of a perfect gas}
\end{tabular}
\end{table}
}
%

\subsection{Direct cycle}
\label{sec:directcycle}
\begin{figure}[!!!h]
\centering
\includegraphics[width=13cm]{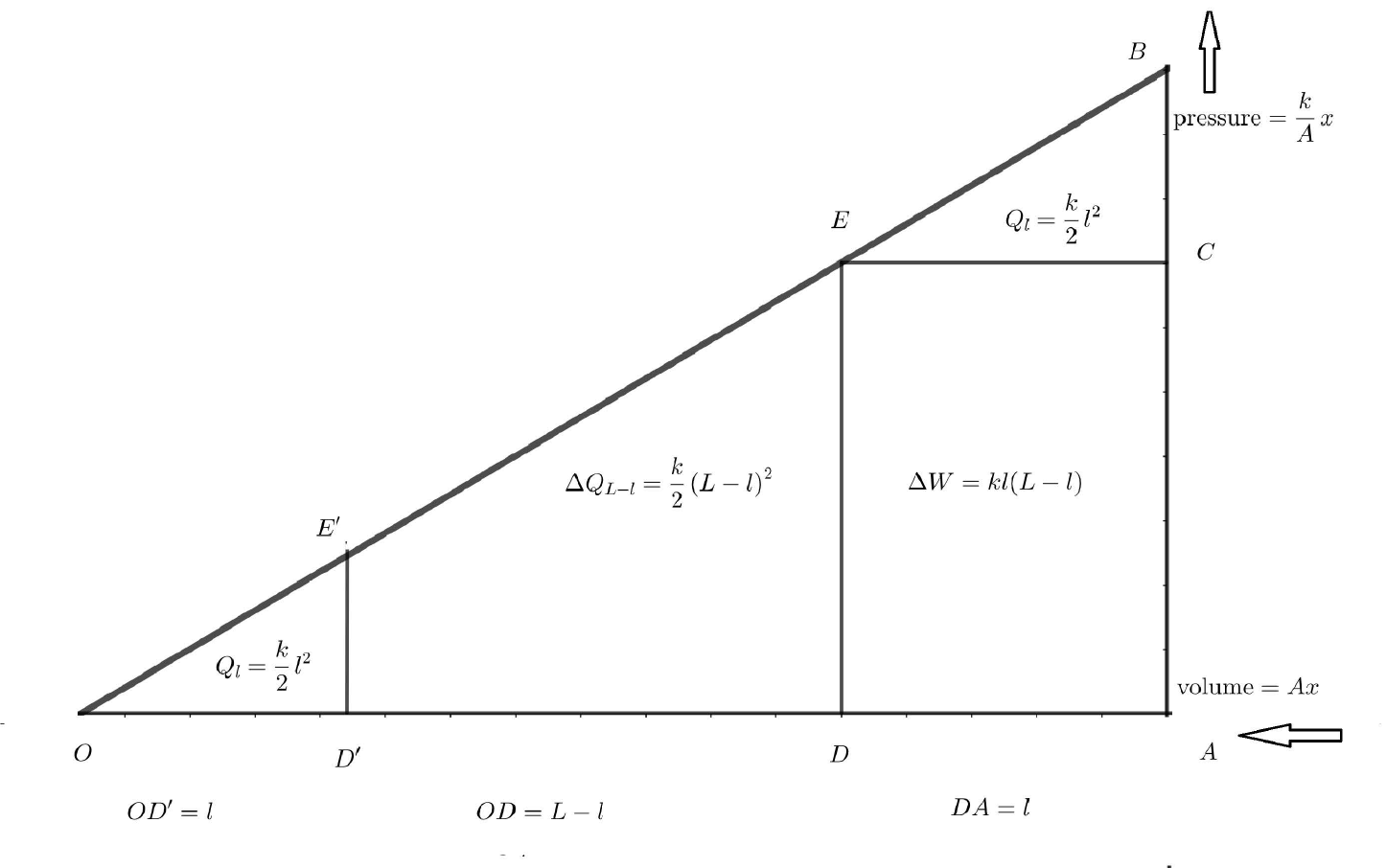}
\caption{The cycle performed by the spring. 
The spring is fully relaxed at $O$, and fully compressed at $A$.
The abscissa is the volume  $Ax$; the ordinate is the pressure on the spring $P_x=\frac{k}{A}x$. The cycle is: $OB$ -- $BE^{\prime}$ -- $E^{\prime}O$.} 
\label{fig:fig3}
\end{figure}

The direct cycle is graphically represented in Figure~\ref{fig:fig3}; 
the spring undergoes three processes:  
``isothermal" compression, ``adiabatic" expansion  and ``isothermal" expansion, in the sense in Table~\ref{tab:dictionary}.

At the beginning of the direct cycle:  valves $\nu_1$ and $\nu_2$ are closed;  $\cal{H}$ is full with air, $\cal{L}$ is empty; the spring is fully relaxed. 

\subsubsection*{Process 1. 
``Isotherm" compression of the spring  ({\boldmath{$OB$}})}
 Valve $\nu_2$ is closed.  Valve $\nu_1$ is opened,  and air flows from $\cal{H}$ into the syringe, compressing the spring. When the spring is entirely compressed, the operator closes $\nu_1$.

{\sl Graph $\left(V, P\right)$.}
The process starts at $O$, where the spring is fully relaxed (length is $L$), and ends at $B$, where it is fully compressed (length $0$). 
{\bf State at {\boldmath{$B$}}:} the length of the spring is $0$ (fully compressed),  and the pressure in the spring is $P_B=\frac{k}{A}L$.

{\sl Description of the process.}  
The process is  isotherm  in the sense of Table~\ref{tab:dictionary}. 
The process is quasi-static in the sense of Table~\ref{tab:dictionary}, so that  the work  done on the spring by the air (transferred heat) is fully converted in potential energy of the spring (internal energy). Then:
\begin{equation}
\Delta W^{\mbox{\tiny(iso)}}_{\mbox{\tiny done on spring}}=
\frac{k}{2}L^2=\mbox{area $\triangle OBA$}
\label{eq:energiatotal}
\end{equation}

\subsubsection*{Process 2.  ``Adiabatic" expansion  of the spring ({\boldmath{$BE^{\prime}$}})} 
Valves $\nu_1$ and $\nu_2$ are closed. 
The operator places a peg holding the piston, to prevent the expansion of the spring. Valve $\nu_2$ is opened, and 
air flows from the syringe  into 
$\cal{L}$.\ignorespaces 
\footnote{The plastic tubes in Figure~\ref{fig:prototype} are parts of the reservoirs. The expansion of air in the tubes and in the  reservoirs should not be of concern, like processes inside the reservoirs in Carnot's perfect reversible engine. 
The only important process occurring inside  the reservoirs is the following: in Carnot's engines there is an actual cooling down of the hot reservoir, and an actual heating up of the cold reservoir, as the cycles go on, which is taken care of by considering  the reservoirs to be ``big enough"; the un-thermal dynamic is ``finite", and  the reservoirs eventually achieve equal pressure.}
 When some quantity  
of the initial volume of air  supplied by $\cal{H}$  is let  into $\cal{L}$ (say, equal to the quantity in the volume $Al$), the operator closes valve $\nu_2$.
Both  valves are now closed.  The operator removes the peg, and the spring  expands,  doing work on the air.  

{\sl Graph $\left(V, P\right)$.} The process starts at $B$, and ends at $E^{\prime}$. 
{\bf State at {\boldmath{$B$}}:}  
the spring is held by the peg, so that the length of the spring is $0$;
but the spring has an extra energy, 
$\frac{k}{2}
l^2=\mbox{area of ${\triangle}EBC$}$, so that it will expand from $B$ to $O$, spending an energy equal to $\frac{k}{2}L^2+\frac{k}{2}l^2$. 	
{\bf State at {\boldmath{$E^{\prime}$}}:} 
the length of the spring is $\frac{AD^{\prime}}{A}=L-l$, the pressure on the spring is $P_{E^{\prime}}=
\frac{k}{A}l$. 

{\sl Description of the process.}   External work is done on the spring (by holding it with a peg),  so that it does not expand, while air leaves the syringe. When valve $\nu_2$ is closed and the peg  removed,    the spring expands, doing work on the air. The air is initially at a lower pressure, $\frac{k}{A}\left(L-l\right)$, so that the spring  expands, until its pressure is equal to the pressure of  the air;  this happens, when the spring is at  state $E^{\prime}$, having expanded the length $AD^{\prime}=L-l$ (but it still has to loose the  extra energy, $\frac{k}{2}l^2=\mbox{area of 
${\triangle}OD^{\prime}E^{\prime}$}$). There is no transfer of air to both  $\cal{L}$ and $\cal{H}$: the valves are closed;  then the process is adiabatic in the sense of Table~\ref{tab:dictionary}. 
 The process also is quasi-static in the sense of Table~\ref{tab:dictionary},
 so that  the spring does work, using its potential energy (internal energy):
\begin{equation}
\Delta W^{\mbox{\tiny(ad)}}_{\mbox{\tiny done by spring}}=
 \frac{k}{2}\left(L-l\right)^2
\equiv
\frac{k}{2}L^2-\frac{k}{2}l^2-kl\left(L-l\right)
\end{equation}

\subsubsection*{Process 3. ``Isotherm" expansion of the spring  ({\boldmath{$E^{\prime}O$}})} 
Valve $\nu_1$ remains closed. Valve $\nu_2$ is  opened.
The remaining air flows from the syringe to $\cal{L}$, emptying the syringe.
When the syringe is empty, and the spring entirely relaxed, the operator closes  $\nu_2$. 

{\sl Graph $\left(V, P\right)$.}  The process starts at $E^{\prime}$, then   the spring expands, and ends at $O$. {\bf State at  {\boldmath{$O$}}:} the length of spring is $A0=L$ (fully relaxed),  the pressure on the spring is $P_O=0$.

{\sl Description of the process.} 
Valve $\nu_2$ is open, so that the springs expands a length $l$, sending air into $\cal{L}$;  then the process is isotherm in the sense of Table~\ref{tab:dictionary}.  The process also is quasi-static in the sense of Table~\ref{tab:dictionary},
 so that  the spring does work, using its potential energy (internal energy):
\begin{equation}
\Delta W^{\mbox{\tiny(iso)}}_{\mbox{\tiny done by spring}}=\frac{k}{2}l^2= \mbox{(area of ${\triangle}OE^{\prime}D^{\prime}$)}
\label{eq:isothermexpansion}
\end{equation}

At the end of the direct cycle:
the spring is fully relaxed; valves $\nu_1$ and $\nu_2$ are closed; the syringe is empty.  $\cal{L}$ contains a quantity of air equal to the quantity in  $V=AL$.  $\cal{H}$ looses a quantity of air equal to the quantity in  $V=AL$. 
\subsection{Reversed cycle}
\label{sec:reversedcycle}

At the beginning of the reversed cycle, the engine is
as it is at the end of the direct cycle. The operator  opens  valve $\nu_2$, and 
pulls the piston  quasi-statically (in the sense of  Table~\ref{tab:dictionary}), until the air in $\cal{L}$
is back into the syringe. Then the operator closes valve $\nu_2$, opens valve $\nu_1$, and lets the spring expand,  pushing the air back into  
$\cal{H}$.\ignorespaces 
\footnote{The reversed cycle is not essential to the paper. The interesting point is the balance of energy.
The energy available to the operator is the useful work: 
$\Delta W=$ area of rectangle $DECA$.
To bring the spring to state $E^{\prime}$, the operator must use the energy
$\Delta W-\mbox{area of ${\triangle}OD^{\prime}E^{\prime}$}$. Then the operator  lets  air into the syringe, quasi-statically, until the spring is at state $E$.   But air and spring are in equilibrium, and
there remains  some air in  $\cal{L}$.
The operator must give an extra push (compressing  the spring by a length $l$), while   the remaining air goes into the syringe (probably, the operator must close valve $\nu_2$ before the push). The total energy   used by the operator is 
$\Delta W-\mbox{area of ${\triangle}OD^{\prime}E^{\prime}$}+
 \mbox{area of ${\triangle}EBC$}=\Delta W$.}


\section{Clausius's 
demonstration  of the two laws of  thermodynamics}
\label{sec:clausiustwooperations}
{\it Carnot's principle} states that the work of a thermal engine consists in the transfer of caloric from a hot to a cold reservoir \cite{Carnot}.  
According to Carnot,   {\em all of} the caloric   taken from the hot reservoir is discarded in the cold reservoir, so that the working substance  does not retain the caloric initially received \cite{DiasUppsala}.\ignorespaces
	 \footnote{Carnot's book was written in the context of the caloric theory, although there are evidences that by 1824 he was skeptical about the caloric.}
 ``Recoverability" of the working substance involves  a conservation law, the conservation of caloric, $\oint dQ=0$ 
 \cite{DiasUppsala,Klein1974}.\ignorespaces
\footnote{The adiabatic processes between the isothermal processes guarantee that  exchanges of heat occurs only between reservoirs \cite{DiasUppsala}.}

After James Prescott Joule's experiments on the equivalence of work  and heat, Clausius \cite{Clausius1850} corrected Carnot's principle:  part of the heat taken from the hot reservoir is transformed in work, and the remaining part is discarded. He recognized then two operations in a thermal engine:\ignorespaces
\footnote{
In Carnot's cycle, call $T_h$ the temperature of the hot reservoir, $T_l$ the temperature of the cold reservoir, and  $V_1$ and $V_2$ respectively the initial and the  final volumes in the isothermal expansion. 
The  work done on the gas during the isothermal compression of the gas is the heat transferred from the hot reservoir (up to omitted constants):
 $\left|\Delta Q_{\mbox{\tiny hot}}\right|=T_h\ln\left(V_2/V_1\right)$. 
The ``transferred heat" is (up to omitted constants):
 $\left|\Delta Q_{\mbox{\tiny cold}}\right|=T_l\ln\left(V_2/V_1\right)$.
The ``consumed" heat is their  difference:
$\left|\Delta Q_{\mbox{\tiny consumed}}\right|\equiv\left|\Delta W_{\mbox{\tiny useful}}\right|
=\left(T_h-T_l\right)\ln\left(V_2/V_1\right)$. 
}
\begin{enumerate}
\item
Operation 1: ``consumption" (Clausius's word)  of heat. According to {\it Joule's principle}, the ``consumed" heat is equal to the work (up to a constant that transforms units of heat in units of work). 
\item
Operation 2: transfer of heat from a hot to a cold reservoir.
This is {\it Carnot's principle} applied to the remaining heat.
 \end{enumerate}

The first operation is ruled by {\it Joule's principle}, and leads to the first law of 
thermodynamics.\ignorespaces
\footnote{Clausius considered only perfect gases. He  shows  that Joule's principle implies that there exists a  function $U$ such that $dU= \bar{d}Q+\bar{d}W$.}
The second operation is ruled by {\it Carnot's theorem}, and leads to the second law of thermodynamics.

Carnot's theorem states  that two engines  that differ only in the choice of the working substance produce the same quantity of work. In  Clausius's version of the theorem 
 \cite{Clausius1850}, the engines produce the same quantity of work;  then work generated by the direct engine is used to operate the reversed engine. It is also supposed that they transfer different quantities of heat. If the engine that transfers the bigger quantity of heat is made to operate in the reversed cycle,  the coupling of the engines results in a  quantity of heat  from the cold to the hot reservoir without external work.
 To forbid this result, Clausius postulates a new qualitative law  \cite[p.~134]{Clausius1850}:  ``[heat] always shows a tendency to equalize temperature differences and therefore to pass from {\em hotter} to {\em colder} bodies".
 
 In 1854, Clausius   \cite{Clausius1854}  found the quantitative, analytic  expression of the new law.
  Clausius's reasoning may be interpreted as follows. If in each  single cycle, the engine ``consumes" the same amount of heat, and  ``transfers"  the same amount of heat, then  the two operations are  ``canceled" (in some sense) at the end of the cycle; otherwise,   the generated work would not be sufficient to restore the hot reservoir
    upon reversal of the engine, and conversely,  the hot reservoir would eventually not deliver enough heat to produce  the fixed amount of work. The calculation leading to the analytic law is lengthy \cite{Dias1995}. It suffices to present the result:
for a reversible engine, ``cancellation" is:
\begin{equation}
+|Q_W|f(T_h)-|Q_t|
\left(f(T_l)-f(T_h)\right)=0;
\nonumber
\end{equation}
$T_h$ is the higher temperature; $T_l$ is the lower temperature; $Q_W$ is  the ``consumed" heat;  $Q_t$  is the transferred heat; $f\left(T\right)$ is an unknown function of the temperature.   
Clausius called $+|Q_W|f(T_h)$ the ``equivalence value" of the transformation of heat in work at  temperature $T_h$ (``consumption"),  and 
$-|Q_t|
\left[f(T_l)-f(T_h)\right]$  the ``equivalence value" of the transfer of heat from $T_h$ to $T_l$. 
``Cancellation" means that the ``values of equivalence" of the two operations cancel each other. 
 Noting that $|Q_W|+|Q_t|=|Q_h|$ is the heat taken from the hot reservoir, ``cancellation" is:
 \begin{equation}
+|Q_h|f\left(T_h\right)-|Q_t|f\left(T_l\right)=0; \nonumber
\end{equation}
Clausius redefines $f\equiv 1/T$; then   $T$ now designates an unknown function of the temperature. Finally,  ``cancellation" is: 
$$\sum\frac{\Delta Q_i}{T_i}=0.$$ 
In 1865, Clausius \cite{Clausius1865} made $T$ equal to the temperature, in analogy with perfect gases, and created the word `entropy'.  
\section{The two laws of the un-thermal dynamic engine}
\label{sec:twolaws}
The  un-thermal dynamic engine performs the two operations recognized by Clausius in thermal engines:
\begin{enumerate}
\item
Production of a quantity of work equal to (Figure~\ref{fig:fig3}):  
$$\Delta W= \mbox{area of parallelogram $ACED$}\equiv 
\mbox{area $\triangle\!\!\!\nabla{ACED}$}= kl\left(L-l\right).$$
\item
Transfer of air from $\cal{H}$ to $\cal{L}$:  the amount of air in the volume $V=AL$. 
\end{enumerate}

\subsection{The first law}
\label{sec:law1}
From the cycle (Section~\ref{sec:directcycle}): 
\begin{eqnarray*}
\mbox{energy input}&\equiv& 
\Delta U=\Delta W^{\mbox{\tiny(iso)}}_{\mbox{\tiny done on spring}}=\frac{k}{2}L^2;
\\
\mbox{wasted energy}&\equiv&\Delta Q=
\Delta W^{\mbox{\tiny(ad)}}_{\mbox{\tiny done by spring}}
+ 
\Delta W^{\mbox{\tiny(iso)}}_{\mbox{\tiny done by spring}}
\\
&=&
\frac{k}{2}L^2- kl\left(L-l\right);
\\
\mbox{``consumed" energy}&=&
\Delta U-\Delta Q=kl\left(L-l\right)
\end{eqnarray*}

Joule's principle is:
$$
\mbox{\bf  available work}= \mbox{\bf ``consumed" energy}$$
or
$$
\Delta W=\Delta U-\Delta Q.
$$
The  the first law of the un-thermal dynamic engine is:
{\boldmath
\begin{equation}
\Delta U=\Delta W+\Delta Q.
\label{eq:firstlaw}
\end{equation}}

The law expresses a condition on the possibility of resetting (``recoverability" of) the spring (working substance) at the end of a cycle, so that 
the spring recovers  its  initial state (volume and pressure). 
In fact, 
if during a cycle the spring is damaged, loosing its elasticity, it is not possible to store in it the energy $\Delta U=(k/2)L^2$, in order to 
perform identical cycles.\ignorespaces
\footnote{Analogously, Carnot discards the use  of steam as working substance, because  the steam is condensed (or ``destroyed", in his words) at the end of a cycle; each cycle must be performed {\em using the same [working] substance}, demands Carnot.}
``Recoverability" of the spring (working substance) --- hence  the possibility of ``resetting" the cycle --- is possible, if and only if
there exists a function 
 $\Delta U=\Delta W+\Delta Q$ that remains  constant in each  closed cycle:
{\boldmath{
$$\sum_{\mbox{\tiny\bf closed cycle}}\Delta U=0.
$$}}
\subsection{The second law}
\label{sec:law2}
Following Clausius, some sort of ``cancellation" must be  postulated. The amount of air used in a cycle is fixed by the maximum compression of the spring ($L$, in the case), and it is equal to the  amount in the volume $Al+A\left(L-l\right)=AL$. If there is a leak of air in the engine, the reversed cycle cannot restore an amount of air equal to the amount delivered by $\cal{H}$ at the beginning of a cycle; this amount can only be restored by pumping new air into $\cal{H}$, i.e., by ``burning new fuel".
In fact, if the leak occurs between $\cal{H}$ and the syringe, 
in order to compress the spring by $L$, $\cal{H}$ has to deliver more air than the amount in $V=AL$; if the leak occurs between the syringe and $\cal{L}$, at the end of a cycle $\cal{L}$ contains less air than the amount in $AL$; or leaks occur in both places.
 The condition on the ``recoverability" of $\cal{H}$ is that at the end of each  cycle, the quantity of air in $\cal{L}$ is exactly equal to the total amount of air  delivered by $\cal{H}$ at the 
beginning of the cycle.  From Figure~\ref{fig:fig3}:
\begin{eqnarray*}
OA&=&V=AL
=
\mbox{volume of air delivered by ${\cal{H}}$} 
\\
DA=OD^{\prime}&=&
v_1=Al=\mbox{volume of air sent to ${\cal{L}}$ to lower the pressure}
\\
OD=D^{\prime}A&=&v_2= A\left(L-l\right)
=
\mbox{volume of air sent to ${\cal{L}}$ to close the cycle}
\end{eqnarray*}
 Then, ``cancellation"  is, indifferently,  any of the equations:
\begin{subequations}
\begin{alignat}{2}
AL-Al-A\left(L-l\right)=0
 \label{eq:unthermalcancellationd}
\\
V-v_1-v_2=0
 \label{eq:unthermalcancellatione}
\end{alignat}
\end{subequations}
With few algebraic manipulation,   
Equation~\ref{eq:unthermalcancellationd}
is written in a form similar to the  second law of thermal engines: 
\begin{equation}
\frac{\frac{k}{2}L^2}{\frac{k}{A}L}
-
\frac{\frac{k}{2}l^2}{\frac{k}{A}l}
-
\frac{\frac{k}{2}\left(L-l\right)^2}
{\frac{k}{A}\left(L-l\right)}\equiv0,\nonumber
\label{eq:unthermalvalueofequiv1}
\end{equation}
which is: 
\begin{equation}
\frac{\Delta U}{P_L}-\frac{\Delta Q_l}{P_l}-
\frac{\Delta Q_{L-l}}{P_{L-l}}\equiv 0
\label{eq:unthermalvalueofequiv2}.
\footnote{Recalling that 
$\Delta U=\Delta W+\Delta Q_{L-l}+\Delta Q_l$, 
Equation~\ref{eq:unthermalvalueofequiv2} can be re-written:
 $$
+\frac{\Delta W}{P_L}-
\Delta Q_l\left[\frac{1}{P_l}-\frac{1}{P_L}\right]-
\Delta Q_{L-l}\left[\frac{1}{P_{L-l}}-\frac{1}{P_L}\right]\equiv0.$$
$+\frac{\Delta W}{P_L}$ is  the analogue of  the ``equivalence value" of the transformation of heat in  work at the  temperature $T$. In the un-thermal dynamic engine, it is the work delivered  by the engine at the pressure $P_L$.

 $-\Delta Q_l\left[\frac{1}{P_l}-\frac{1}{P_L}\right]$ and $-\Delta Q_{L-l}\left[\frac{1}{P_{L-l}}-\frac{1}{P_L}\right]$ are  analogues of the  ``equivalence value"  of the transfer of heat from the temperature  $T_h$ to the temperature  $T_l$.   In the un-thermal dynamic engine, the first expression is associated with the transfer of air at pressure $P_L$ to air at pressure  $P_l$ (hence, the spring can do ``isothermal" work equal to  $\Delta Q_l$);  
 the second is associated with with the transfer of air at pressure $P_L$ to air at pressure $P_{L-l}$ (hence, the spring can do ``adiabatic" work  equal to $\Delta Q_{L-l}$).
}
\end{equation}
With the sign convention that work done on the spring is $+$, and work done by the spring is $-$, and using $Q_1$, $Q_2$ and $Q_3$ to denote the work respectively  involved in the three processes, Equation~\ref{eq:unthermalvalueofequiv2} can be written as the algebraic sum:
\begin{equation}
\frac{\Delta Q_1}{P_1}+\frac{\Delta Q_{2}}{P_2}+\frac{\Delta Q_3}{P_3}\equiv0,\qquad\mbox{or}\qquad {\sum_{j=1}^3}
\frac{\Delta Q_j}{P_j}\equiv0.
\label{eq:unthermalentropylaw}
\end{equation}

If the engine is not airtight, the leaked air causes a loss of pressure (denoted by $P_{\mbox{\tiny leaked}}$), which entails a loss of work (denoted by $\Delta Q_{\mbox{\tiny leaked}}<0$, negative, according to the sign convention.
 To accomplish ``cancellation", the lost work has to be taken into consideration:
\begin{equation}
\left({\sum_{j=1}^3}
\frac{\Delta Q_j} 
{P_j}\right)+\frac{\Delta Q_{\mbox{\tiny leaked}}}{P_{\mbox{\tiny leaked}}}=0\qquad\mbox{or}\qquad{\sum_{j=1}^3}
\frac{\Delta Q_j} 
{P_j}=-\frac{\Delta Q_{\mbox{\tiny leaked}}}{P_{\mbox{\tiny leaked}}}>0.
\label{eq:noneqentropy}
\end{equation}

The second law of the un-thermal dynamic  engine  states: 
\begin{itemize}
\item 
If the engine is airtight,  the air initially delivered by $\cal{H}$ can be used cycle after cycle. 
 Equation~\ref{eq:unthermalentropylaw} means that  there  exists a function --- the entropy ($\Delta{\cal{S}}$) --- so defined:
{\boldmath{
\begin{equation}
\Delta{\cal{S}}=\frac{\Delta Q}{P}, \quad\mbox{such that}\quad\sum_{\mbox{\tiny \bf closed cycle}}\Delta{\cal{S}}=0;
\label{eq:entropia}
\end{equation}
}}
i.e., the ``recoverability'' of the reservoirs is expressed by {\bf the conservation of 	{\boldmath{${\cal{S}}$}}  in any cycle}. 
\item
If the engine is not airtight, Equation~\ref{eq:noneqentropy} means that there exists a quantity of heat, $\Delta Q_{\mbox{\tiny leaked}}$, for which there  is not  a  definite  analytic expression, so that there exists an analytically undefined $\Delta{\cal{S}}$, such that
{\boldmath{$$
\Delta{\cal{S}}>\sum_{\mbox{\tiny \bf closed cycle}}\frac{\Delta Q}{P}>0.
$$}}
\end{itemize}
\section{The epistemic meaning of entropy}
\label{sec:epistemicentropy}
In ``Limitations of the Second Law of Thermodynamics",  a very appropriately named chapter in his {\em Theory of Heat} \cite{MaxwellHeat}, James Clerk Maxwell argues that the second law is macroscopic. To substantiate the claim, he created a Demon, later named after him, a being able to tamper with molecules in a gas. This creature uses the fluctuation of velocities of the molecules in a gas, initially in  thermal equilibrium, to bring the gas back to a state out of equilibrium. The gas occupies the two halves of a container, separated by a wall with a trap door; by operating the door, the Demon lets fast molecule  go to one of the halves, and slow molecules to the other half. If the two halves have  initially the same temperature, the result of the operation is to make their temperatures unequal. 

In 1929, Leo Szilard \cite{Szilard}  proposed three thought experiments to explain why a little intelligent being could not operate. The point is that, in order to act, i.e., either open or close the trap door, the Demon  needs to know if there is a molecule moving towards the trap door, and if it is fast or slow. In other words, he has to acquire information on the fluctuations of velocities in the gas. Entropy is spent in acquiring this information; the desideratum is that the balance between the reduction of entropy made by the Demon, and the entropy increase in obtaining information, should be enough to save the second law. 

In the 1950s, Leon Brillouin \cite{BrillouinI,BrillouinII,Brillouinbook} 
attempted to show how acquisition of information accounted for the necessary entropy increase.
 In the first paper in the series, he illustrates the relation between information and experiments.
Accordingly, the Demon has to use a flash-light to be able to see the molecules, absorbing photons, which increases the  entropy of the Demon.
 Dennis Gabor \cite{Gabor}  proposed a thought experiment similar to Szilard's first experiment, but instead of a creature that saw molecules, he imagined a beam of light confined to one of the halves of the cylinder; to confine the beam, goes the argument,  demands a more intense beam of light, and a consequent entropy cost. As appealing as they are, these arguments   rely  on  circumstances of observation; for instance,
 Brillouin's interpretation of the cost of information  relies on circumstances of experimentation rather than on general principles \cite[p.~136]{BrillouinI}:
\begin{quote}
{\footnotesize The physicist in his laboratory is no better off than the [D]emon. Every observation he makes is at the expense of the negentropy [negative entropy] of his surroundings. He needs batteries, power supply, compressed gases, etc., all of which represent sources of negentropy. The physicist also needs light in his laboratory to be able to read ammeters or other instruments.}
\end{quote}

\subsection{The three-tapes reversible computer}
\label{sec:thereversiblecomputer}
New categories on which the Demon's action could be explained came from   computation. In 1961, Rolf Landauer \cite{Landauer1961} argued that the process of computation was irreversible, due to the irreversibility of the logical operations ``and"  and  ``or": according to the truth-table, the operation ``and" ($\wedge$) is a three-to-one map that brings three states into the state ``false" ($0$), and cannot be reversed; the operation ``or" ($\vee$) is a three-to-one map that brings three states into the state ``true" ($1$), and cannot be reversed.\ignorespaces
\footnote{$p$ and $q$ are sentences that can be 
either true ($1$) or false ($0$).
The  proposition  ``$p$ and $q$" ($p\wedge q$) is true, if  $p$ and $q$ are both true, i.e., if the state  $1\wedge1$ occurs; it is false, when any one of the remaining three states occur:  
$1\wedge0$ or $0\wedge1$ or $0\wedge0$.
The  proposition  ``$p$ or $q$" ($p\vee q$) is false, if $p$ and $q$ are both false, i.e., if the state  $0\wedge0$ occurs; it is true, when any one of the remaining three states occur:  
$1\wedge1$ or $1\wedge0$ or $0\wedge1$.
}

Bennett's theorem states that for any one-tape irreversible Turing machine there is a three-tapes reversible Turing machine  \cite{Bennett1973}.
The computer operates in three stages
 (Table~\ref{tab:threetapecomputer}). 
{
\renewcommand{\baselinestretch}{0.8}
\begin{table}[!h]
\caption{The reversible computation  (adapted from \cite{Bennett1988}). The position of the read/write heads is indicated by  ``$\_$". }
\label{tab:threetapecomputer}
\centering
{\footnotesize
\begin{tabular}{l|l|l|l}
\\
\hline
\textbf{stage and the task}
&
\textbf{work tape} 
& 
\textbf{history tape} 
& 
\textbf{output tape}
\\
\textbf{performed in it}
&&
\textbf{or garbage tape}
&
\\
\hline\hline
forward:
&
\underline{\phantom{X}}{INPUT} 
&
\underline{\phantom{X}}
&
\underline{\phantom{X}}
\\
(1) do the computation
&
{COMPUT}\underline{A}{TION}
& 
{COMPUT}\underline{\phantom{X}}
&
\underline{\phantom{X}}
\\
(2) copy the computation
&
\underline{\phantom{X}}{OUTPUT}
&
{COMPUTATION}\underline{\phantom{X}}
&
\underline{\phantom{X}}
\\
\hline
copy-the-output 
&
\underline{\phantom{X}}{OUTPUT}
&
{COMPUTATION}\underline{\phantom{X}}
&
\underline{\phantom{X}}
\\
&
{OUT}\underline{{P}}{UT}
&
{COMPUTATION}\underline{\phantom{X}}
&
{OUT}\underline{\phantom{X}}
\\
&
\underline{\phantom{X}}OUTPUT
&
{COMPUTATION}\underline{\phantom{X}}
&
\underline{\phantom{X}}{OUTPUT}
\\
\hline
reverse or cleanup:
&
\underline{\phantom{X}}{OUTPUT}
&
{COMPUTATION}\underline{\phantom{X}}
&
\underline{\phantom{X}}{OUTPUT}
\\
(1) follow back the  history
&
{COMPUT}\underline{A}{TION}
& 
{COMPUT}\underline{\phantom{X}}
&
\underline{\phantom{X}}{OUTPUT}
\\
(2) retrieve the input
&
\underline{\phantom{X}}{INPUT}
&
\underline{\phantom{X}}
&
\underline{\phantom{X}}{OUTPUT}
\\
\hline
\end{tabular}
}
\end{table}
	}

 In the {\sl forward stage}, the calculation takes place on the work tape, and is also copied in the history or garbage tape.\ignorespaces
\footnote{
The operation of making a history tape is equivalent to  copying information on a blank tape, which is reversible:  to copy information  on a blank tape is a one-to-one map, the tautology, $A\equiv A$.
Landauer \cite[p.~28]{Landauer1991}  illustrates the point with physical examples.}
In the {\sl copy-the-output stage}, the output is copied on the output tape. In the {\sl reverse or cleanup stage}, the history tape is used to reverse the operation, and it is erased as the computation goes on.  At the end, the history tape is erased, the input is retrieved, and the output remains recorded.

According to the three-tapes computer, reversibility is thus associated with  retrieval of information, and irreversibility with erasure of information, not with its acquisition. Although irreversibility is associated with logic, 
it is physical \cite{Landauer1991}. This is ``Landauer's principle"; it is also stated as the principle that  erasure of information entails an increase of entropy  ``in the environment or in the degrees of freedom that do not carry information" \cite[p.~502]{Bennett2003}. This can be illustrated by Bennett's interpretation of Szilard's first example \cite{Bennett1982}. 
After being confined to one of the halves of a cylinder, a single  molecule is brought back to the equilibrium situation in which it occupies the whole cylinder; this is done by moving the partition separating the halves toward the empty side, so that the molecule can now move in  the whole cylinder, which  erases the information that tells in which side the molecule was initially confined. The Demon's mind has two  possible phase space states, either a side  or the other side of the cylinder;  when information is acquired,  the Demon's mind is in one of two of its possible  states, but when information is erased,  the two possible phase space states are ``compressed" in one single state (the ``uninformed" one, the whole cylinder); in Bennett's interpretation  ``all the work obtained [\dots] must be converted into heat again in order to compress the demon's mind back to its standard [uniformed] state" \cite[p.~237]{Bennett1982}.  
\section{The thermodynamic entropy is a measure of  erased information}
\label{sec:unthermalcomputer}
Bennett's  theorem discloses  a general
principle of computation;  it discloses 
a condition inherent in algorithms that  arises from logic --- hence is universal ---  not from circumstances.
When applying the theorem  to, say,  the un-thermal dynamic engine, an analogy 
is made between the algorithm of calculation, and 
the operations performed by the engine; these are   the two operations performed by 
the  un-thermal dynamic engine, or, what is the same, the processes undergone by the engine in a cycle (Section~\ref{sec:twolaws}). In 
order to identify the nature of the erased information,  identify the garbage tape. To exemplify, consider Szilard's second example.\ignorespaces
\footnote{
A cylinder closed at the top has its bottom made of a membrane  permeable to only a kind of molecule; the cylinder can slide inside another cylinder closed at the bottom, but whose top is a membrane permeable to a different kind of molecule. At the beginning, the inner cylinder is entirely in the outer cylinder, and is full with the two kinds of molecules. As the inner cylinder slides up, the bottom membrane keeps only one kind of molecule  inside it, and lets the other kind go into the outer cylinder. When the cylinder has slid its full length, the membranes are in contact; then the membranes are exchanged. Moving the inner cylinder down, the molecules are again mixed.}
In this example, two different kinds of molecules initially mixed can be separated by the action of two membranes, respectively permeable to a kind of molecule, but not to the other kind; the example is discussed by Harvey S. Leff and Andrew F. Rex, who  ``[\dots] {\em interpret} the membrane actions as constituting ``measurements [\dots]"  \cite[p.~161]{LeffRexMembrane}. The membranes together are the ``history or garbage tape": they ``inform" the  kind of molecule contained in each cylinder;  by exchanging membranes, 
it is possible to mix the system again, performing the reverse operation.  

\subsection{Analogy between the three-tapes computer and the un-thermal dynamic engine}
\label{sec:theunthermalcomputer}
Parts of the un-thermal dynamic engines  can be interpreted  as the tapes in the three-tapes computer:
\begin{quote}
		{\sl WORK TAPE:  
$\cal{H}$ $+$  syringe $+$ air in the syringe.}
The tape records:  the input (the amount of air in initial volume of air, $V=AL$), the direct cycle and the work.
\\
{\sl GARBAGE TAPE: $\cal{L}$.}
The tape records: 
 the amount of air that  leaves the syringe in the two expansion processes, respectively the amounts in 
   $v_1=Al$ and in $v_2=A\left(L-l\right)$.
$\cal{L}$ is qualified as a ``garbage tape":  the un-thermal dynamic engine is reversible, only if  the air stored in $\cal{L}$ 
accounts for all the air involved in the operation 
(Section~\ref{sec:law2}).
\\
{\sl OUTPUT TAPE: spring.}
The tape records the output: the useful work $\Delta W$ is recorded as elastic potential energy.
\end{quote}
Similarly to the three-tapes computer, the 
un-thermal dynamic engine  performs three stages (Table~\ref{tab:analogywithcomputation}):

{\renewcommand{\baselinestretch}{0.8}
\begin{table}[!h]
\caption{Computation stages in the un-thermal dynamic engine. They are similar to the stages in Table~\ref{tab:threetapecomputer}}
\label{tab:analogywithcomputation}
\centering
{\footnotesize
\begin{tabular}{l|l|l|l}
\\
\hline
\textbf{stage}
&
\textbf{work tape:}
& 
\textbf{garbage tape:}
& 
\textbf{output tape:}
\\
&
\textbf{{\boldmath{${\cal{H}}$}},
syringe, air in the syringe}
&
\textbf{{\boldmath{$\bf{\cal{L}}$}}}
&
\textbf{spring}
\\
\hline\hline
forward
&
\underline{\phantom{X}}{INPUT: $V=AL=v_1+v_2$} 
&
\underline{\phantom{X}}
&
\underline{\phantom{X}}
\\
&
perform the direct cycle\underline{\phantom{X}}
& 
$v_1+$\underline{\phantom{X}}
&
\underline{\phantom{X}}
\\
&
\underline{\phantom{X}}{OUTPUT: $\Delta W$}
&
$v_1+v_2$\underline{\phantom{X}}
&
\underline{\phantom{X}}
\\
\hline
copy output
&
\underline{\phantom{X}}OUTPUT: $\Delta W$
&
{$v_1+v_2$}\underline{\phantom{X}}
&
\underline{\phantom{X}}
\\
&
$\Delta W$\underline{$=$}
&
$v_1+v_2$\underline{\phantom{X}}
&
$\Delta W$\underline{\phantom{X}}
\\
&
\underline{\phantom{X}}OUTPUT: $\Delta W$
&
$v_1+v_2$\underline{\phantom{X}}
&
\underline{\phantom{X}}OUTPUT: $\Delta W$
\\
\hline
reverse
&
\underline{\phantom{X}}OUTPUT: $\Delta W$
&
{$v_1+v_2$}\underline{\phantom{X}}
&
\underline{\phantom{X}}OUTPUT: $\Delta W$
\\
&
perform the reversed cycle\underline{\phantom{X}}
&
$v_1+$\underline{\phantom{X}}
&
\underline{\phantom{X}}OUTPUT: $\Delta W$
\\
&
\underline{\phantom{X}}INPUT: $V=AL=v_1+v_2$ 
&
\underline{\phantom{X}}
&
\underline{\phantom{X}}OUTPUT: $\Delta W$
\\
\hline
\end{tabular}
}
\end{table}}
\begin{quote}
{\sl FORWARD STAGE.} The input, the direct cycle and the output are recorded in the work tape, in this order. The amount of air in $v_1$ and $v_2$ is recorded in the garbage tape. Nothing happens in the output tape.
\\
{\sl COPY-OUTPUT STAGE.} The work tape and the garbage tape are frozen. The output is copied  into the output tape.
\\
{\sl REVERSE STAGE.} The output tape is frozen. The output, the reversed cycle and the input are recorded in the work tape, in this order. The garbage tape is erased as the work tape is being recorded. 
\end{quote}

What is stored in $\cal{L}$ is ``information" in the sense that it contains what is needed to restore the condition of $\cal{H}$: this information is measured by  quantity of air, 
therefore it  is physical, in agreement with Landauer's principle; the erased information can also be measured in unity of volume:
volume occupied by the air at the end of the ``isothermal" expansion of the spring (equal to $AL$) minus 
 the volume occupied by the air 
restored to the syringe in the reversed cycle. It is also ``information" in  Szilard's epistemic sense of knowledge that can be used to bring a system that has reached  thermodynamic equilibrium, back to a situation out of thermodynamic equilibrium, so that work can again be done:  the reverse cycle brings the spring  from the relaxed (un-thermodynamic) equilibrium state into a compressed, out of (un-thermodynamic) equilibrium state.
Entropy is straightforwardly associated with measure of erased information or missing information. 
\section{Final remarks}
\label{sec:conclusion}
The interpretation of entropy as (measure of) ``missing information" was proposed by Edwin T. Jaynes \cite{Jaynes1,Jaynes2}, in the context of  the justification of the ensembles in equilibrium statistical mechanics. Jaynes proposed that statistical mechanics was an instance of the ``maximum entropy principle" (MEP). This is a method of inductive predictions about the state of, for example, a physical system: the most honest method of induction is a method that does not make non warranted assumptions about the system. Taking ``entropy" as the measure of  the amount of information needed to specify the exact state of the system, the most honest inductive method is the one that maximizes the entropy. Cardozo Dias and Abner Shimony \cite{DiasShimony} placed the  MEP in Rudolf Carnap's \cite{Carnap} classification of inductive methods; it is identified with an inductive method discarded by Carnap, because it makes strong assumptions about the system. 
In statistical mechanics, this is the assumption that the molecules move independently of each other, even if they collide and interact, which is one of the problems on the foundations of statistical mechanics that the ergodic theory intends to solve \cite{LebowitzPenrose}; in the theory, equilibrium statistical mechanics results from phase-space properties of dynamic systems. 

The epistemic meaning of entropy as missing or erased information is an additional hypothesis in physics that asks for proof. The un-thermal dynamic engine is an attempt to overcome  this situation at the macroscopic level. 

I hope to have succeeded in answering a question  Abner Shimony  posed me,  regretfully a too belated answer.
\begin{flushright}
{\em To Abner Shimony in gratitude, and with affection.
\\
Saudades
}
\end{flushright} 
\section*{Acknowledgments}
Mr. Herc\'ilio P. C\'ordova is a former graduate student in the ``Mestrado Profissional de Ensino de F\'isica". I showed him 
Figure~\ref{fig:1}, and asked: ``is this thing feasible?" His answer was the prototype, and the video. To 
my colleague and friend Professor Carlos Eduardo Aguiar, many thanks for his comments, and for believing in my work, which means a lot to me.

\end{document}